\documentstyle[twoside,fleqn,espcrc2,epsfig]{article}

\def\lsim{\mathrel{\rlap{\lower4pt\hbox{\hskip1pt$\sim$}}
    \raise1pt\hbox{$<$}}}         
\def\gsim{\mathrel{\rlap{\lower4pt\hbox{\hskip1pt$\sim$}}
    \raise1pt\hbox{$>$}}}         

\def\frac#1#2{{{#1}\over {#2}}}

\def\smallfrac#1#2{\hbox{${{#1}\over {#2}}$}}

\def\GeV{{\mbox{\rm GeV}}}

\def\MS{\hbox{$\overline{\rm MS}$}}

\newcommand{\AmS}{{\protect\the\textfont2
  A\kern-.1667em\lower.5ex\hbox{M}\kern-.125emS}}

\begin{document}
\begin{titlepage}
\setcounter{page}{0}
\textheight 206mm                 
\topmargin 2mm
\begin{flushright}
{\tt hep-ph/9610268}\\
{DFTT 61/96}\\
{Edinburgh 96/25}\\
\end{flushright}
\vskip 12pt
\begin{center}
{\bf DOUBLE ASYMPTOTIC SCALING '96} 
\vskip 24pt
{Stefano Forte}\\
\vskip 12pt
{\em INFN, Sezione di Torino,}\\ 
{\em via P.~Giuria 1, I-10125 Torino, Italy }\\ 
\vskip 18pt
{and}
\vskip 18pt
{Richard D. Ball\footnote[1]{Royal Society University Research Fellow}}\\
\vskip 12pt
{\em Department of Physics and Astronomy}\\
{\em University of Edinburgh, Edinburgh EH9 3JZ, Scotland}\\
\vskip 36pt
\end{center}
{\narrower\narrower
\centerline{\bf Abstract}
\medskip\noindent
We review recent HERA data on the structure function $F_2$ at small $x$ and
large $Q^2$. We show that the salient features of the data are revealed
by comparing them to the double asymptotic scaling behaviour which
$F_2$ is predicted to satisfy in perturbative QCD.\\\smallskip}

\smallskip
\vspace{.3in}
\begin{center}
{Invited talk given at\\\smallskip {\em QCD96}\\ Montpellier, July 1996}\\
\smallskip
{\em to be published in the proceedings}\\
\end{center}
\bigskip
\vfill
\begin{flushleft}
{October 1996}
\end{flushleft}
\end{titlepage}

\title{Double Asymptotic Scaling '96
                                      \thanks{Presented by SF}}

\author{Stefano~Forte\address{INFN,
Sezione di Torino, via P.~Giuria 1, I-10125 Torino, Italy} and
 Richard~D.~Ball\address{Department of Physics and Astronomy,
University of Edinburgh, Edinburgh EH9 3JZ, Scotland}\thanks{Royal
Society University Research Fellow.}}

\begin{abstract}
We review recent HERA data on the structure function $F_2$ at small $x$ and
large $Q^2$. We show that the salient features of the data are revealed
by comparing them to the double asymptotic scaling behaviour which
$F_2$ is predicted to satisfy in perturbative QCD.

\end{abstract}

\maketitle

Double asymptotic scaling of the structure function $F_2(x,Q^2)$ in the 
two variables
$\sigma\equiv\sqrt{\xi\zeta}$ and $\rho\equiv\sqrt{\xi/\zeta}$
(where $\xi\equiv\smallfrac{x_0}{x}$ and
$\zeta\equiv\smallfrac{\alpha_s(Q_0)}{\alpha_s(Q)}$)~\cite{das} is
a simple consequence~\cite{DGPTWZ} of the QCD evolution equations at 
small $x$ and large $Q^2$, and manifests the basic structure of 
asymptotic freedom in
perturbative QCD~\cite{wil}. Because double scaling is spoiled
if the dominant behaviour of $F_2$ at large $\sigma$ and $\rho$ is not
governed by perturbative evolution, plots of $F_2$ as a function of the
scaling variables $\sigma$ and $\rho$ were proposed in ref.~\cite{das}
as a means of testing graphically the validity of the QCD prediction.
Indeed, this analysis showed at a very early stage that,
contrary to widespread expectation, perturbative QCD in 
the leading log $Q^2$ approximation accurately describes 
the behaviour 
\begin{figure}[hb]
\vbox{\vskip-3.2truecm\hskip-0.5truecm
\epsfig{figure=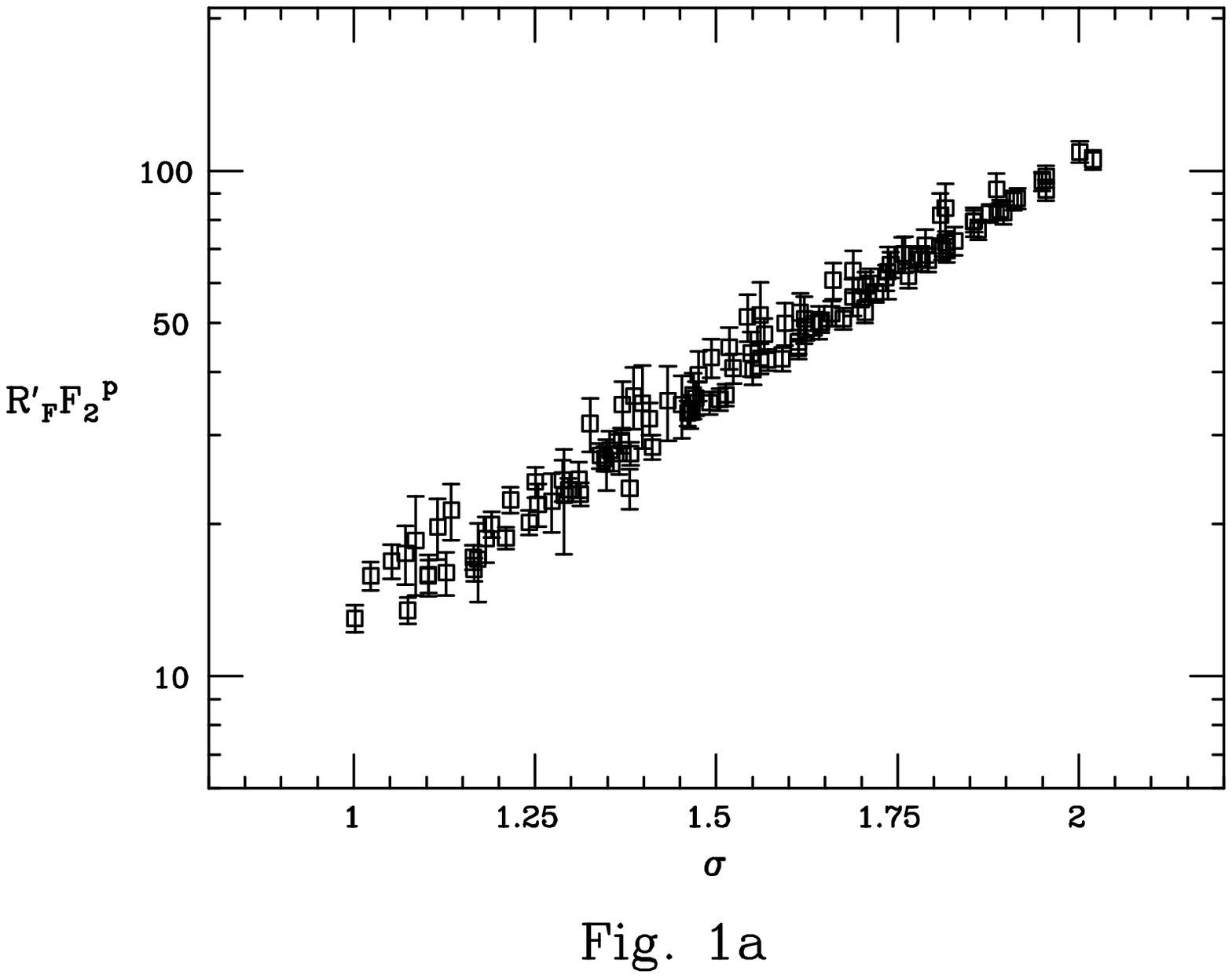,width=8.2truecm}}
\vskip-3.0truecm
\end{figure}
of the structure functions measured at HERA.

While the availability of much more precise data on $F_2$~\cite{hone,zeus}
now justifies a full NLO analysis~\cite{alphas,romasx,romaas} based
on the Altarelli-Parisi equations, whose applicability in this regime
is now established beyond reasonable doubt, many instructive features
of the small $x$, large $Q^2$ behaviour of $F_2$ are apparent from its
double scaling behaviour.

The basic double scaling behaviour of the data is displayed in Fig.~1, which
shows all data on $F_2$ which pass the cuts $\rho,$ $\sigma>1$, $Q^2>5$~GeV$^2$
(so that the leading-order approximation is valid), as determined in the 1994
HERA run by H1~\cite{hone} (a) and  ZEUS~\cite{zeus} (b),
plotted versus $\sigma$ (with $x_0=0.1$ and $Q_0=1$~GeV),
after rescaling of 
the leading one-loop sub-asymptotic behaviour~\cite{das}.
Independence of $\ln F_2$ from $\rho$ ($\rho$ scaling) is demonstrated
by the fact that all data lie on the same line.
\begin{figure}[hb]
\vbox{\vskip-3.2truecm\hskip-0.5truecm
\epsfig{figure=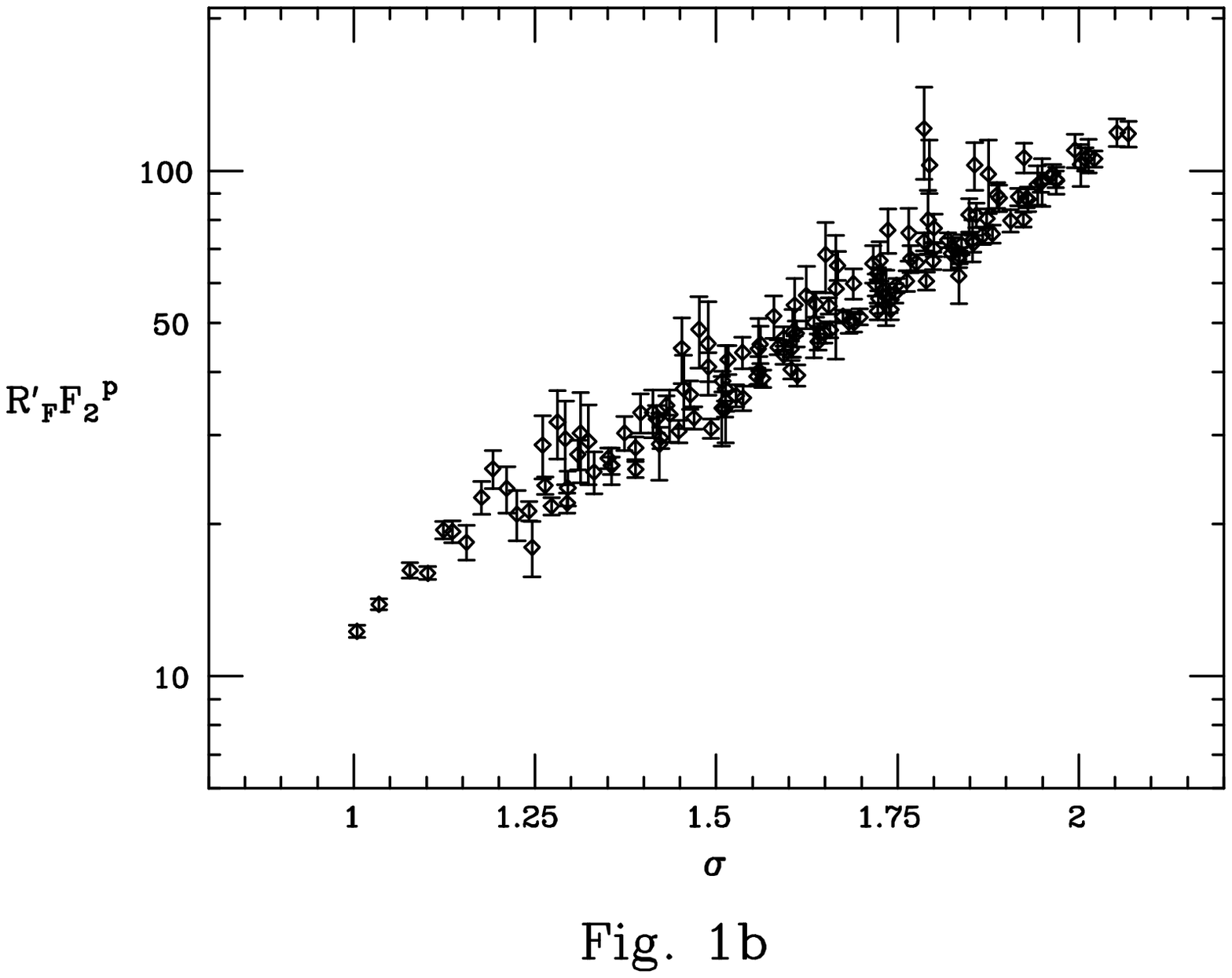,width=8.2truecm}}
\vskip-1.8truecm
\end{figure}

\begin{figure}[t!]
\vbox{\vskip-3.2truecm\hskip-1.0truecm
\epsfig{figure=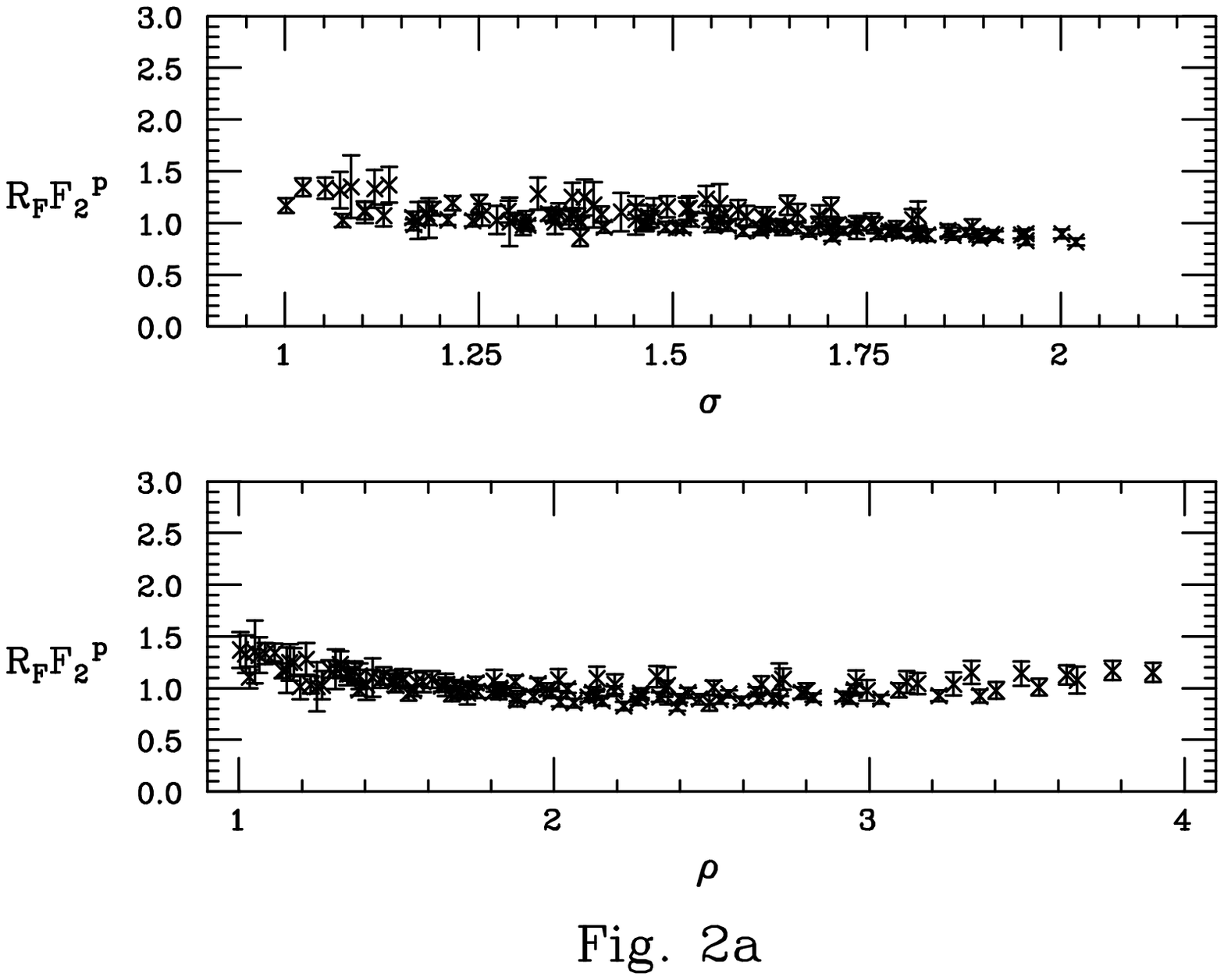,width=8.4truecm}}
\vskip-3.2truecm
\vbox{\vskip-1.8truecm\hskip-1.0truecm
\epsfig{figure=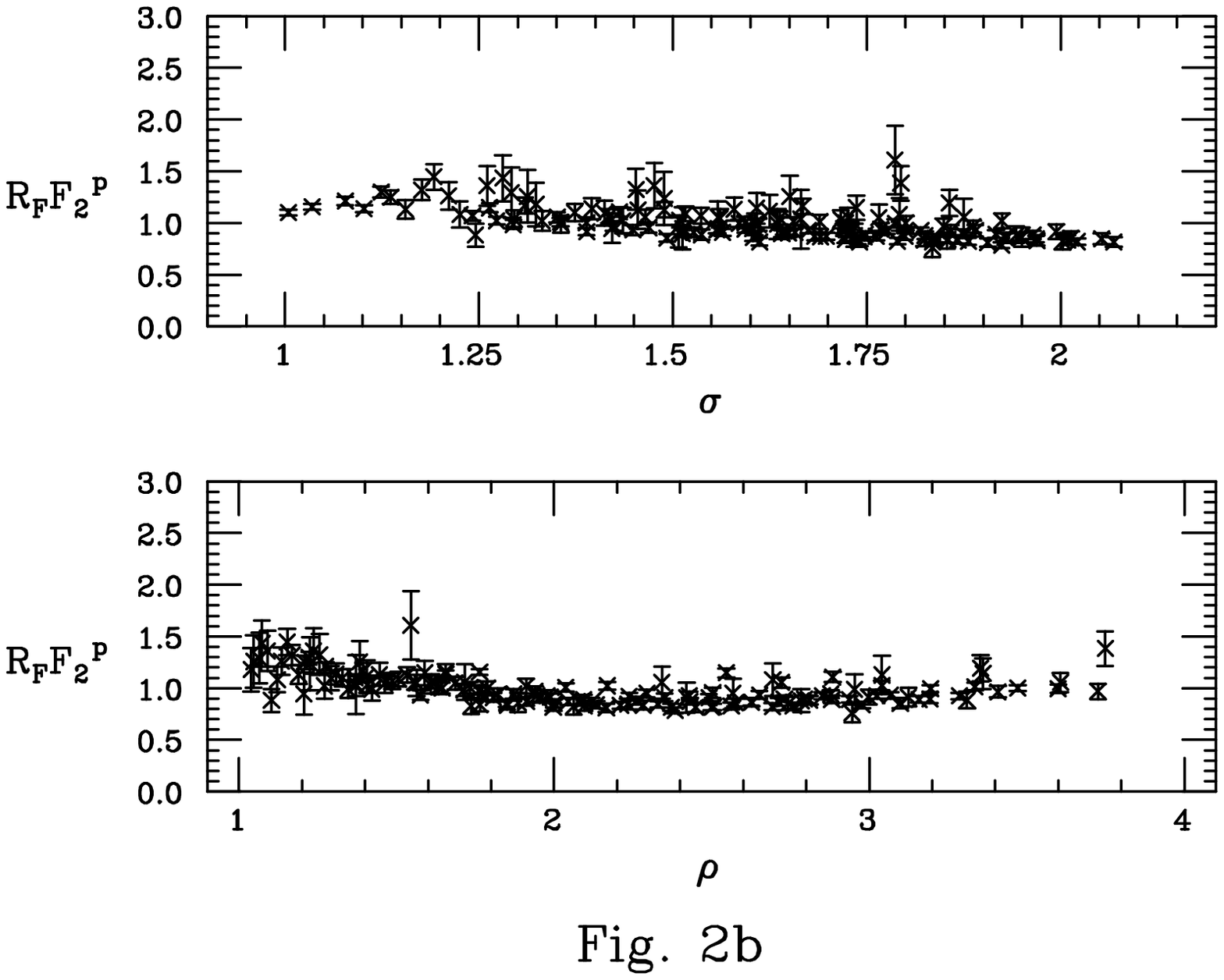,width=8.4truecm}}
\vskip-3.3truecm
\end{figure}

In order to test for $\sigma$ scaling, however, we 
must make sure that the slope $2\gamma$
of this line is that predicted in perturbative QCD,   i.e. that the observed
scaling is not an accidental
consequence of the particular kinematic region covered by the HERA
experiments. Note that the value of $\gamma=\sqrt{12/\beta_0}=6/
\sqrt{33-2 n_f}$ is an absolute,
parameter-free prediction of perturbative QCD. A simple way of doing
so is to scale away
the full asymptotic prediction, i.e. to rescale the data by
the factor
\begin{equation}
R_F=N\rho\sqrt{\sigma}e^{-2\gamma\sigma+\delta \sigma/\rho},
\label{dasbeh}
\end{equation}
with $N$ an  arbitrary normalisation.

\begin{figure}[t!]
\vbox{\vskip-3.2truecm\hskip-1.0truecm
\epsfig{figure=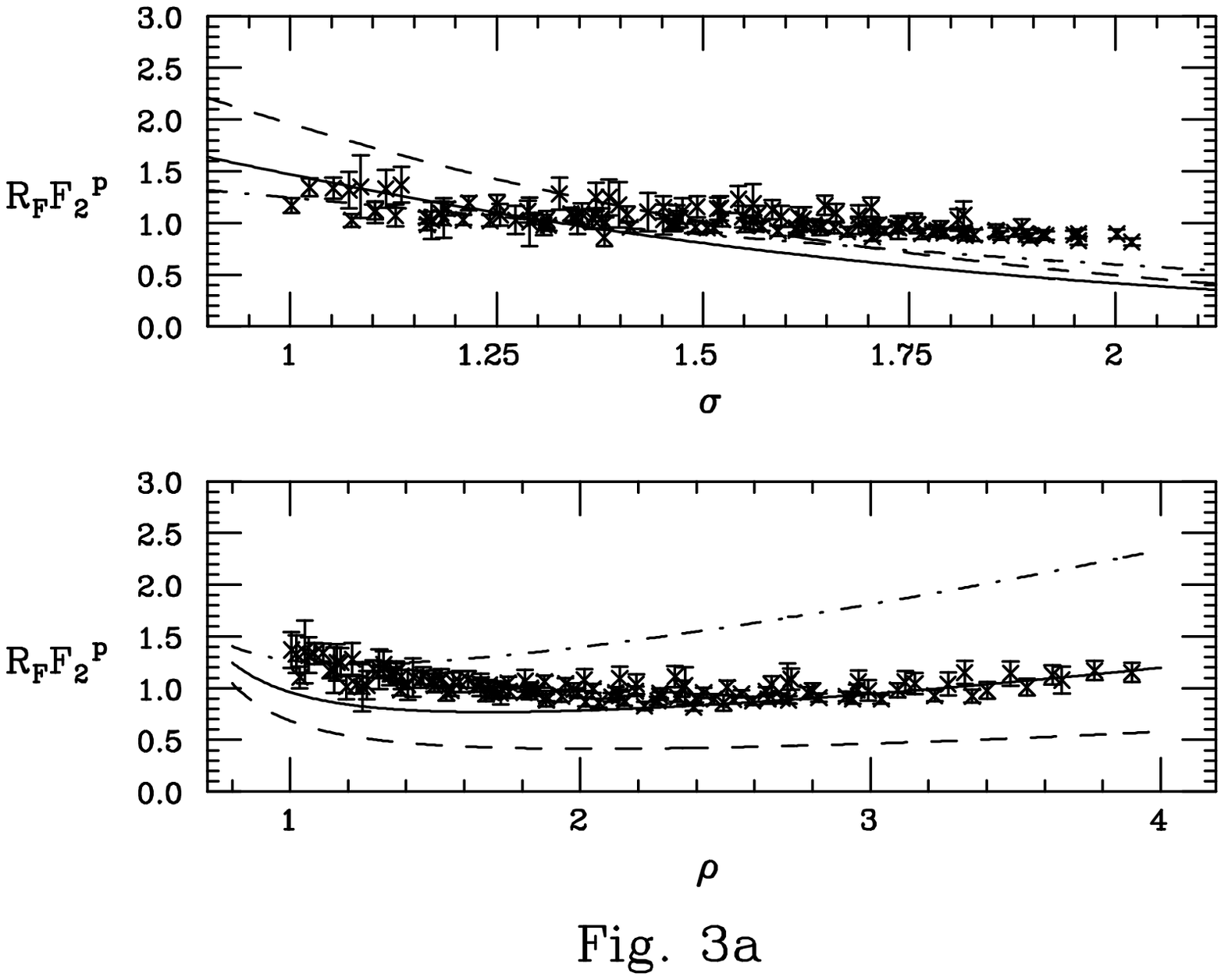,width=8.4truecm}}
\vskip-3.2truecm
\vbox{\vskip-1.8truecm\hskip-1.0truecm
\epsfig{figure=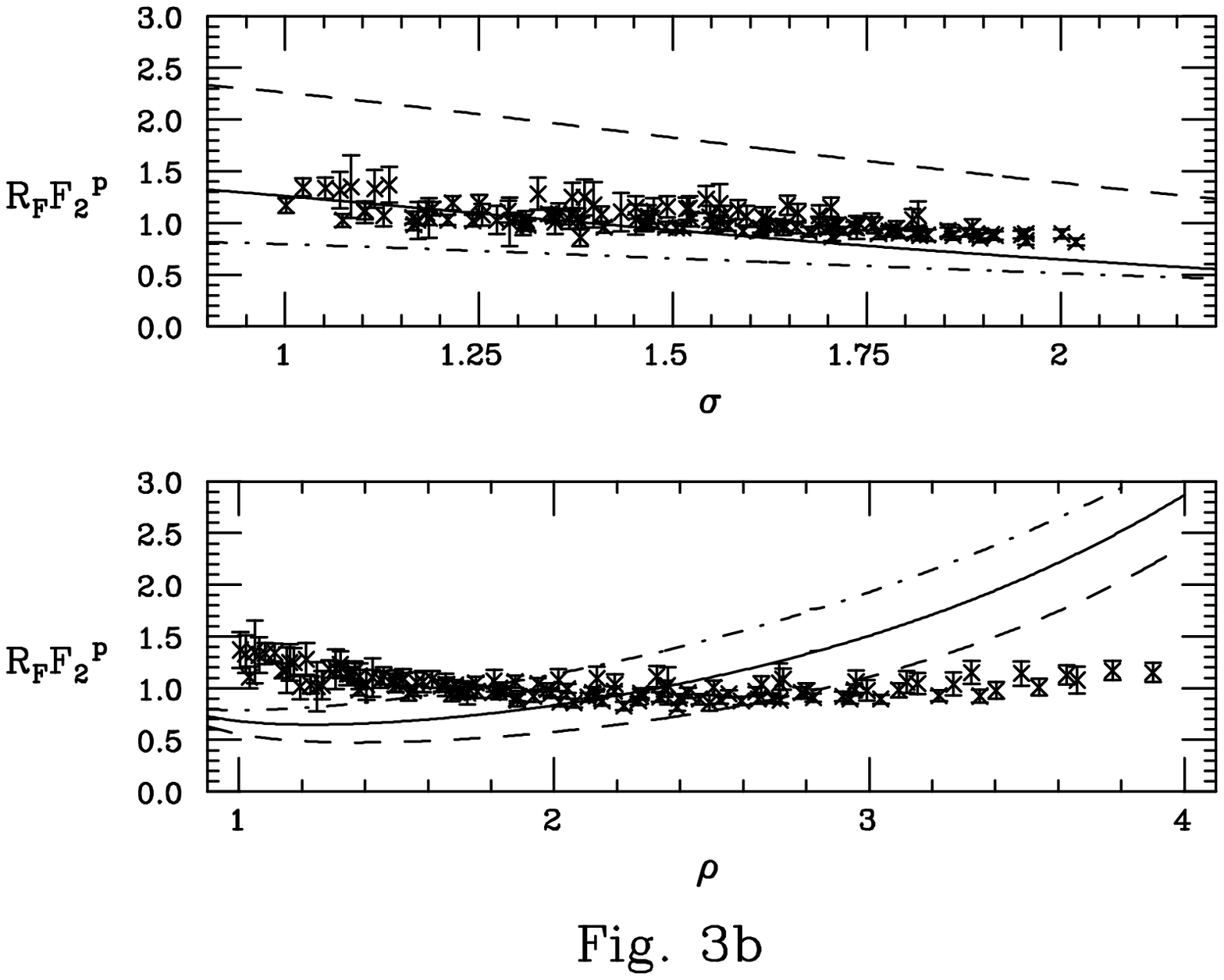,width=8.4truecm}}
\vskip-3.3truecm
\end{figure}

The result is displayed in fig.~2, where, as a cross-check, the dependence
on $\rho$ is also shown explicitly.  Clearly both the H1 (a)
and ZEUS (b) data are in good agreement with the QCD prediction, which
is attained asymptotically at large $\sigma$. The (rescaled) data
are now plotted on a linear scale: as a consequence, because of the
definition of the scaling variables, deviations from the predicted
behaviour which are linear in $x$ or $\log Q^2$ would appear as
exponential scaling violations.

For example, we can immediately exclude that the data behave as a
scale-independent fixed power: in fig.~3 the H1 data are compared to
the function $A x^{-\lambda}$ ($A$ is an arbitrary constant) with
(a) $\lambda=0.08$ and (b) $\lambda=0.35$.\footnote{In this and 
subsequent plots the
three curves correspond to $\rho$=1.1, 2.2, 3.3 on the $\sigma$ plot,
and $\sigma=1,$ 1.5, 2 on the $\rho$ plot (dot-dashed, solid and
dashed curves, respectively), which bracket the HERA kinematic region.}
This is of some theoretical relevance because a (scale independent)
power-like behaviour $F_2\sim x^{-\lambda}$ would correspond to the
dominance of a fixed Regge pole. The value $\lambda=0.08$ is that
corresponding to the ``soft pomeron'' which dominates high-energy
elastic proton-proton scattering~\cite{dola}, while
$\lambda\gsim 0.35$ is supposed to be~\cite{hard} the smallest
value one may obtain from a computation of the pomeron intercept
based on the Lipatov equation~\cite{bfkl} in the presence of
infrared cutoffs. Larger values of $\lambda$ would
lead to yet stronger violations of double scaling.

Of course such a power-like
behaviour might instead only apply at some starting scale $Q_0$
as an initial condition to perturbative evolution. The
leading asymptotic behaviour of $F_2$ then depends on the kinematic
region~\cite{das}:
when $\rho<{\gamma\over\lambda}$ the double scaling rise dominates
over the boundary condition and $F_2$ displays double scaling, while
for larger values of $\rho$ the boundary condition is reproduced,
up to calculable subleading corrections, i.e.~\cite{das}
\begin{equation}
F_2\sim x^{-\lambda} \left({t\over t_0}\right)^{\gamma^2/\lambda-\delta}
+O\left({1\over\sigma}\right).
\label{hardbeh}
\end{equation}
When $\lambda=0.35$ the transition occurs around $\rho\approx 3.5$:
it follows that fig.~3b allows us to exclude any boundary condition
which is more singular than $\lambda\approx 0.35$. Thus, the observation
of double scaling excludes strongly rising boundary conditions~\cite{das}.

\begin{figure}[t!]
\vbox{\vskip-3.2truecm\hskip-1.0truecm
\epsfig{figure=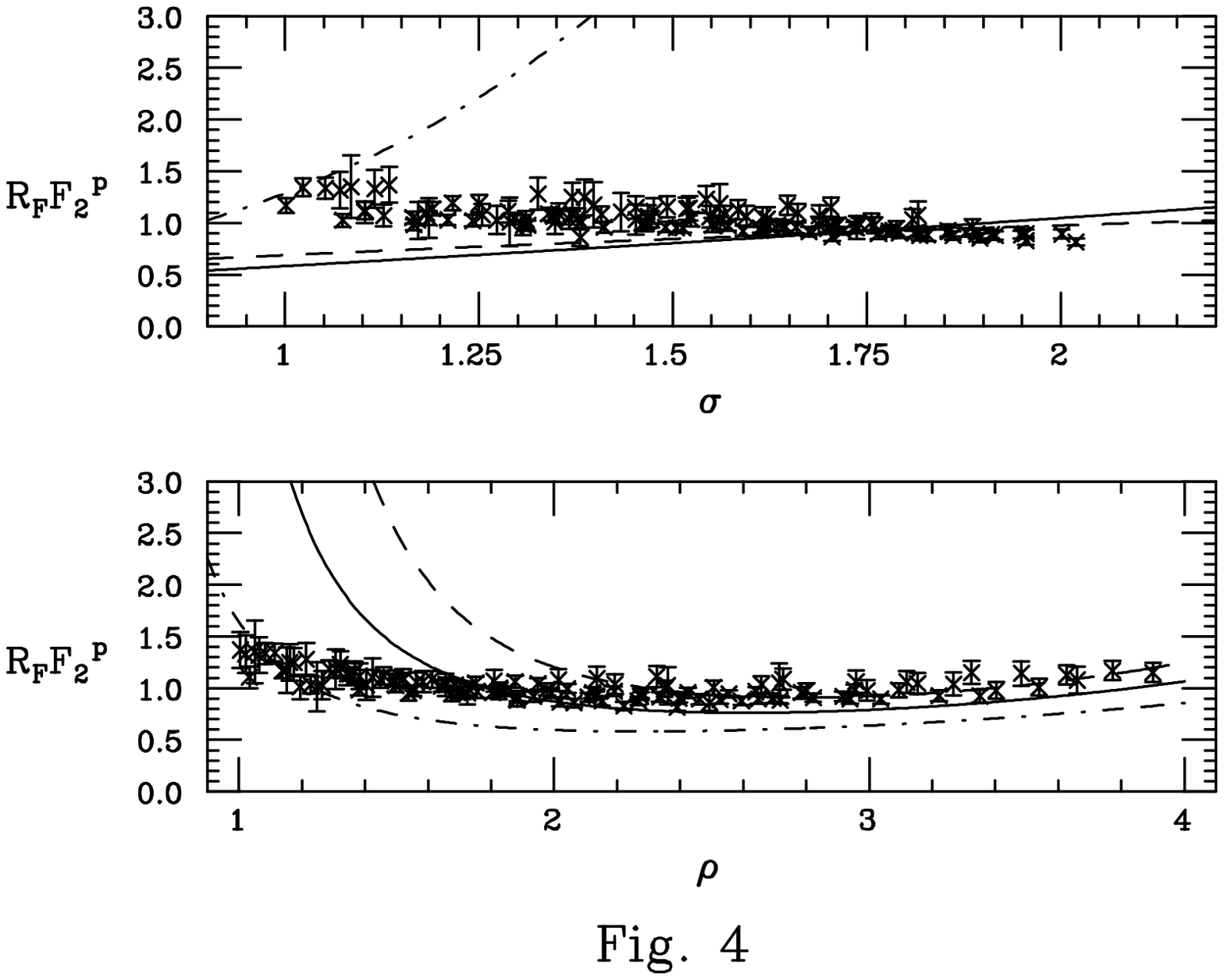,width=8.4truecm}}
\vskip-3.3truecm
\end{figure}
The behaviour eq.~\ref{hardbeh} corresponds essentially to approximating
the Mellin-space anomalous dimensions $\gamma^{ij}(N)$ which
characterise the
evolution of singlet parton distributions with the value which they take
at $N=\lambda$. It is sometimes claimed~\cite{robi} that a 
fixed Regge pole might
dominate the anomalous dimension (rather than the structure function
$F_2$, as assumed in producing the curves of fig.~3). The behaviour 
eq.~\ref{hardbeh} would then be applicable everywhere: of course,
this would mean a breakdown of usual perturbative evolution in the region
$\rho<{\gamma\over\lambda}$ where instead double scaling should hold.
Comparison in fig.~4 of the resulting behaviour (eq.~2 with $\lambda=0.35$)
to the data shows that this scenario is in any case clearly excluded by 
the data essentially because of its large deviation from the perturbative
prediction as $\rho$ decreases, and because the growth with $\sigma$ is 
too steep.

\begin{figure}[t!]
\vbox{\vskip-3.2truecm\hskip-1.0truecm
\epsfig{figure=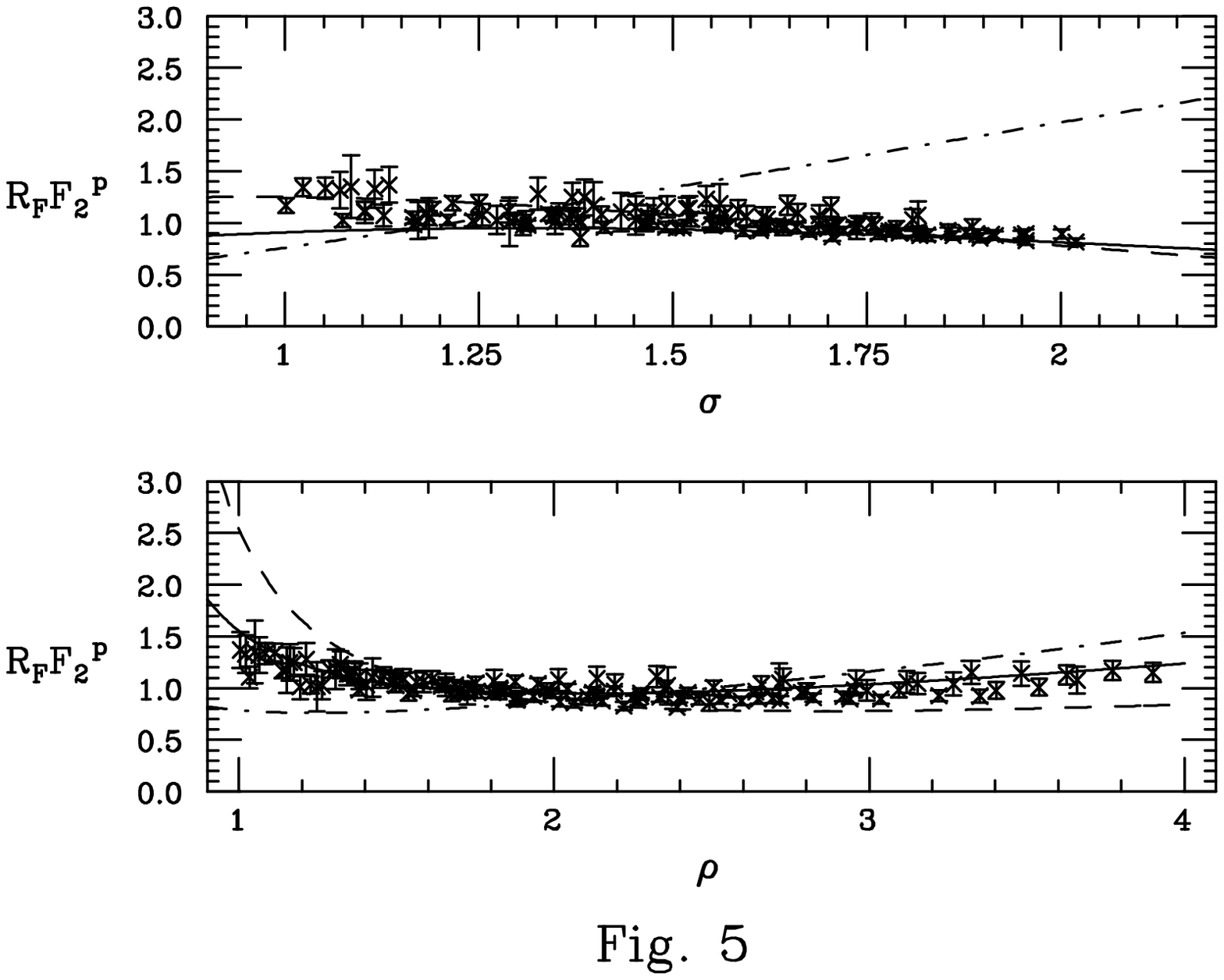,width=8.4truecm}}
\vskip-3.3truecm
\end{figure}
Given that the data seem to exclude most of the commonly proposed 
alternatives to double
scaling, one may wonder whether it is possible to mimic the perturbative
double scaling prediction of eq.~\ref{dasbeh} by a simpler functional form.
In ref.~\cite{buha} a good fit to the H1 data~\cite{hone}
was obtained by assuming $F_2=a+ b\log{Q^2\over Q_0^2}\log{x_0\over x}$,
with $a$ and $b$  free parameters. The corresponding
best-fit~\cite{buha} curves are displayed in fig.~5, which shows that
indeed there is reasonable agreement with double scaling in the HERA region.
The agreement is only local, however: for example at large $\sigma$ and
moderate $\rho$, i.e. at very large $Q^2$ and small $x$ (where
no data are available) the linear growth in $\ln Q^2$ leads to a
discrepancy with double scaling, which corresponds instead to a rise
of $F_2$ weaker than any power of $\ln Q^2$. This
parametrisation thus lacks any predictive power:
the slope of the rise is fitted to the data rather than being predicted, 
and thus cannot be reliably extrapolated
outside the data region. It does however show that, due to the slow
variation of $\xi$ and $\zeta$ with $Q^2$ and $x$, the double scaling
behaviour of eq.~\ref{dasbeh} admits simple local approximations.

Despite its success in describing the data, it is clear from fig.~2
that double scaling is only correct asymptotically. Indeed, a rise
is visible in the $\rho$ plot
at large $\rho$, where $Q^2$ is low and the data get close
to the boundary condition, while the data on the $\sigma$ plot appear to
drop somewhat, except at the largest $\sigma$ values, indicating that
the slope of the rise of $F_2$ is sub-asymptotically smaller than the
predicted one.

We may investigate whether these violations of double
scaling may be understood in a simple way by seeing whether they are reduced
by the inclusion of a two loop correction to the leading double scaling
behaviour. The correction may be computed~\cite{zako} by determining
the NLO correction to the leading singularity of the anomalous dimensions
in $N$ (which has the same location and strength as at leading order), and
deriving the corresponding two loop asymptotic behaviour, which
turns out to be scheme independent.
\begin{figure}[t!]
\vbox{\vskip-3.2truecm\hskip-1.0truecm
\epsfig{figure=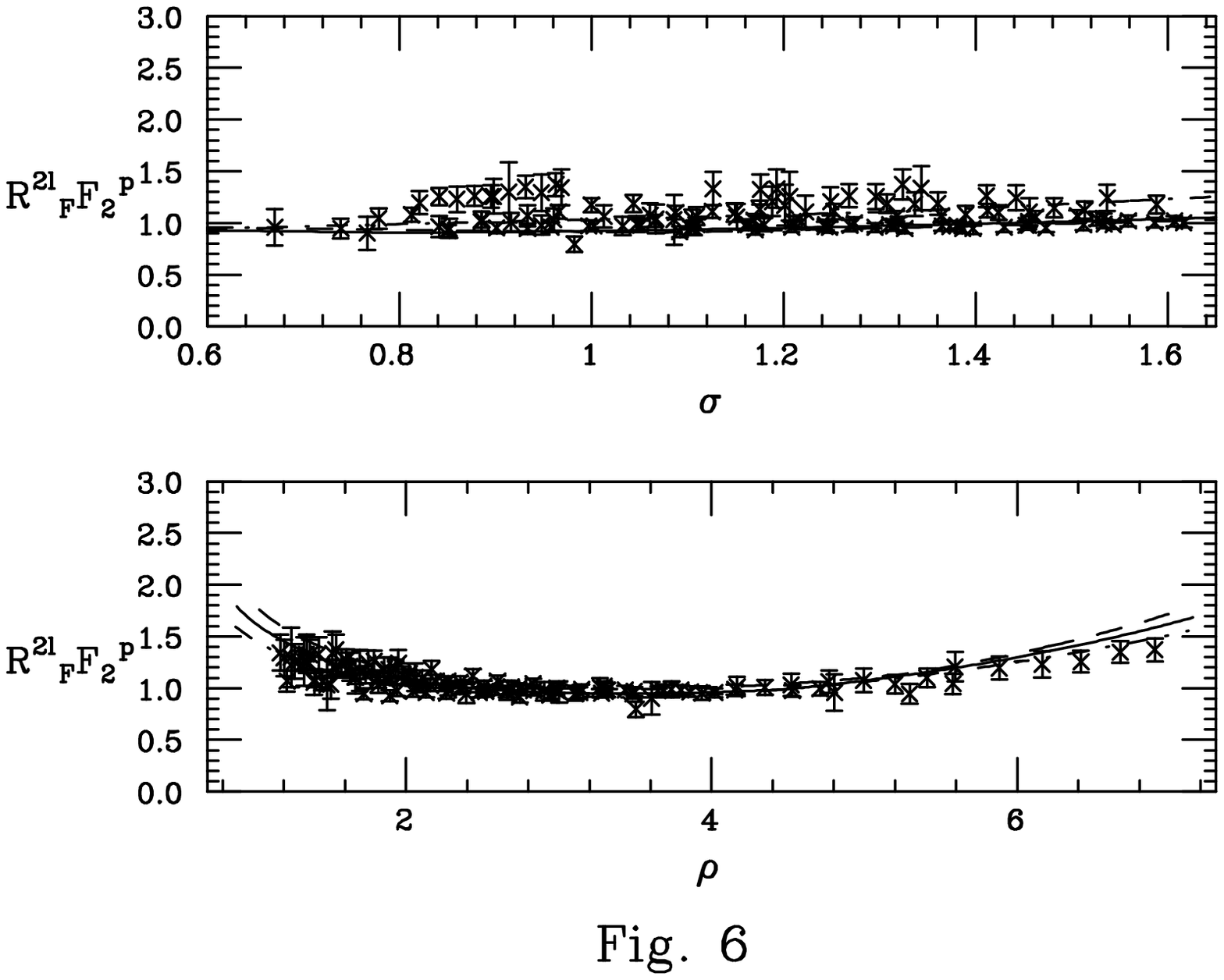,width=8.4truecm}}
\vskip-3.3truecm
\end{figure}
We can then once again test graphically
this two loop correction
by rescaling the data to the two loop asymptotic prediction, i.e.
with the rescaling factor
\begin{eqnarray}
&&R^{2l}_F=\left[1+\smallfrac{\rho}{\gamma}\left(\epsilon\alpha_s(Q^2)-
\epsilon^\prime\alpha_s(Q^2_0)\right)\right]R_F\nonumber\\
&&\quad\epsilon=\smallfrac{1}{\beta_0\pi}\left(\smallfrac{103}{27}+
\smallfrac{3\beta_1}{\beta_0}\right)\\
&&\quad \epsilon^\prime=\epsilon+\smallfrac{78}{\pi\beta_0\gamma^2},\nonumber
\label{2ldas}
\end{eqnarray}
where $\beta_1$ is the
two loop coefficient of the $\beta$ function,
and $R_F$ is given by eq.~\ref{dasbeh}, with $\zeta$ computed
using the two-loop expression for $\alpha_s$.
Fig.~6 shows that indeed the two loop correction eq.~\ref{2ldas}
removes the main systematic violation of double scaling, namely 
the sub-asymptotic fall in the
$\sigma$ plot. The two loop correction also has the effect of
moving upwards the optimal reference scale $Q_0$, to $Q_0=1.8$~GeV  (this
is the main cause of the change of scale in the plots
of fig.~6 compared to the previous one-loop plots).

However, the overall $\rho$ scaling deteriorates somewhat.
Even though part of the $\rho$ scaling violations, in particular those
at large $\rho$, are presumably the result of boundary effects, this
suggests that the part of the two loop anomalous dimension not included
in the determination of the asymptotic behaviour eq.~\ref{2ldas} may
actually be important in most of the HERA region. Otherwise, stated,
this suggests that the summation of logs of $Q^2$ may still have
a leading role over the summation of logs of $1\over x$. This
is consistent with the result of ref.~\cite{romasx}, where
it was shown that the observed perturbative evolution of $F_2$
is most accurately described by two-loop Altarelli-Parisi evolution,
while the inclusion of  the summation of logs of $1\over x$ in the
evolution equation inevitably makes the agreement with the data worse.

This also suggests that a more accurate description of the data requires
a full two loop computation. The results of such a
computation~\cite{romasx,romaas} are also displayed in fig.~6. The curves here
correspond to $\rho$=2, 3, 4 on the $\sigma$ plot,
and $\sigma=1,$ 1.2, 1.4 on the $\rho$ plot (dot-dashed, solid and
dashed curves, respectively).
\begin{figure}[t!]
\vbox{\vskip-3.2truecm\hskip-1.0truecm
\epsfig{figure=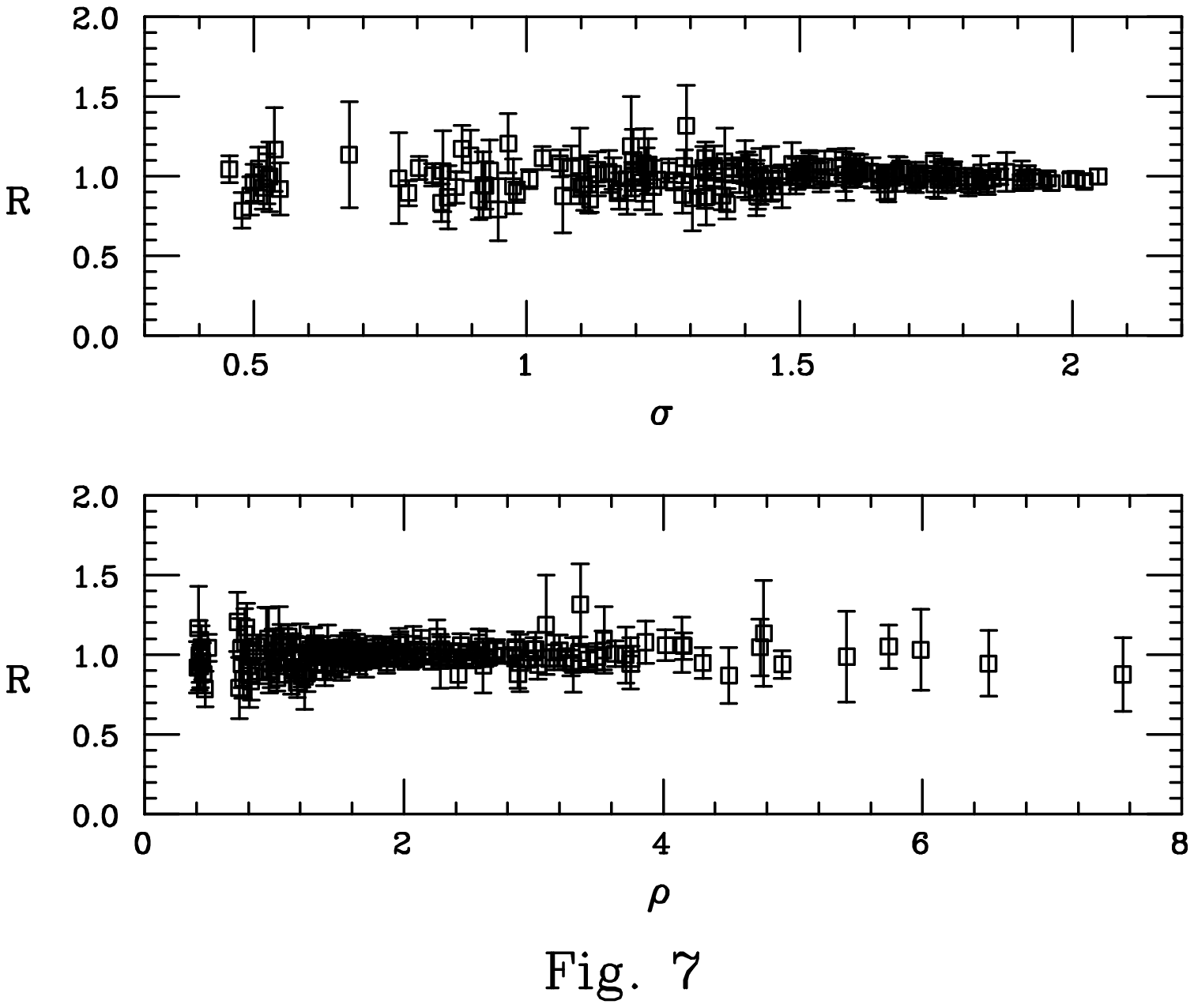,width=8.4truecm}}
\vskip-3.3truecm
\end{figure}
The remarkable accuracy of the
two loop computation is however best shown by plotting the ratio $R$ of the
data to the best fit versus the two scaling variables (fig.~7).
This fit is obtained with only two free parameters, namely the exponents
$\lambda_q$ and $\lambda_g$ which characterise the small
$x$ behaviour $x^{-\lambda_i}$  of the input singlet quark and gluon
distributions (see ref.~\cite{romaas} for details of the fitting procedure).
This is possible because the dominant double scaling behaviour is
largely independent of the details of the input parton
distributions~\cite{das}. In fact, in most of the HERA region any starting
distribution leads to essentially the same behaviour, provided 
only that $\lambda_i$ are not too
large (i.e. $\lambda\lsim 0.35$, as discussed above).

The availability of data at low $x$ and $Q^2$ (i.e large $\rho$ and
large $\sigma$) however, allows now to put more stringent bounds on
the acceptable values of the exponents $\lambda_i$ than the asymptotic
double scaling analysis alone would allow. Indeed,
the full two loop fit~\cite{romaas} leads to the values
$\lambda_q=0.25\pm0.02$ and $\lambda_g=0.09\pm0.09$ at $Q_0=2$\GeV\ 
in the \MS\ scheme: hence the values, say, $\lambda_q=\lambda_g=0.35$
are excluded by several standard deviations,
even though they would only lead to violations of
double scaling at large $\rho\gsim 3.5$.
\begin{figure}[t!]
\vbox{\vskip-3.2truecm\hskip-1.0truecm
\epsfig{figure=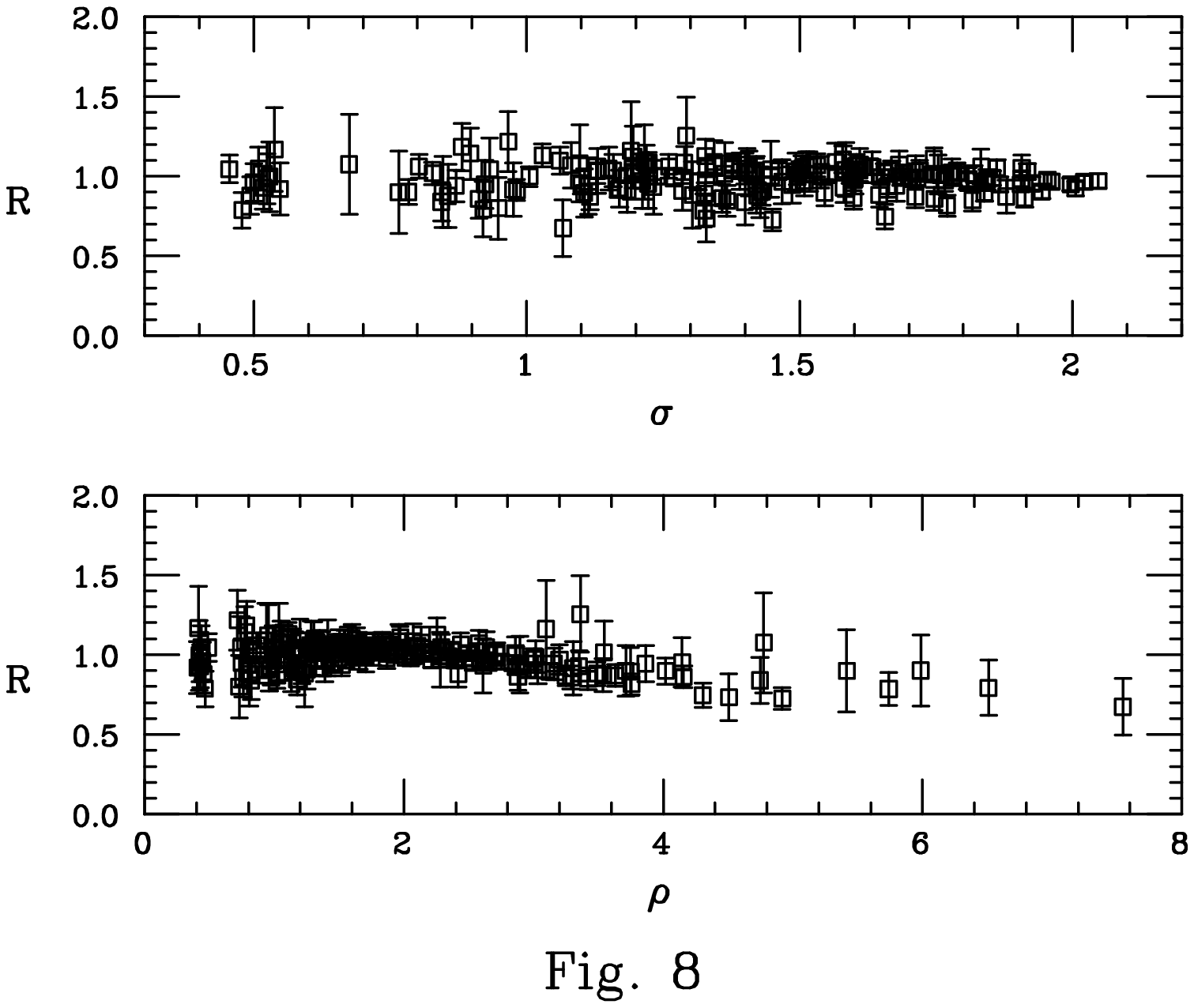,width=8.4truecm}}
\vskip-3.3truecm
\end{figure}
This is apparent in the plot of the data/theory ratio determined with these
values (fig.~8): the excessive rise of the boundary condition
leads the computed $F_2$ to overshoot the data close to the boundary,
and thus to a systematic fall as $\rho$ increases.

The precise values taken by the exponents $\lambda_i$ of course depend on the
particular parameterisation adopted, on the factorization scheme 
and on the value of the starting scale,
and thus do not have any direct physical meaning. It is
nevertheless significant 
that the data require that at low $Q^2$ singlet parton distributions,
and the gluon distribution in particular, do not rise strongly with 
$1\over x$,
in qualitative agreement with expectations based on the soft pomeron of Regge
theory~\cite{dola}.

The description of scaling violations afforded by standard two-loop
Altarelli-Parisi evolution is now accurate enough that scaling violations
due to the summations of logs of $1\over x$ may be excluded
phenomenologically~\cite{romasx}. This, together with the success
of double scaling, which implies the absence of strong nonperturbative
boundary effects which would spoil the perturbatively generated behaviour,
opens the way to very accurate QCD phenomenology at HERA, such
as, for instance, precision determinations of $\alpha_s$~\cite{romaas}.
A complete theoretical understanding of the underlying reasons for 
this happy state of affairs is however still missing, and will certainly 
require going beyond the simple picture of double asymptotic scaling 
summarised here.

{\bf \noindent
Acknowledgements:} We would like to thank S.~Narison for 
insisting that this material be
presented at the conference, and for his hospitality in Montpellier.


\end{document}